\documentclass[twocolumn,secnumarabic,amssymb, nobibnotes, aps, prd]{revtex4}
\usepackage{graphicx}

\begin{document}
\title{\Large\bf Effective Field Theoretical Approach to Black Hole Production}
\vskip 1.5true cm
\author{Sven Bilke~$^{a}$}
\author{Edisher Lipartia~$^{a,b,c}$}
\author{Martin Maul~$^{a}$}
\email{lipartia@thep.lu.se}
\email{maul@thep.lu.se}
\email{sven@thep.lu.se}
\affiliation{\it $^{a}$ Department of Theoretical Physics, Lund University,
S\"olvegatan 14A, S - 223 62 Lund, Sweden}
\affiliation{{\it $^{b}$ Laboratory of Informational Technologies, 
Joint Institute 
for Nuclear Research, 141980, Dubna, Russia}}
\affiliation{{\it $^c$ HEPI, Tbilisi State University, 
University St. 9, 380086, Tbilisi, Georgia}}
\begin{abstract}
A field theoretical description of mini black hole production at TeV energies
is given taking into account the quantization of black holes in discrete
resonances. The unknown quantum gravitational effects are absorbed in 
effective couplings, black hole masses and the Hawking temperature. The
evaporation is described in terms of thermal field theory.\\
\end{abstract}
\maketitle

Recently the possibility to produce mini black 
holes at TeV energies
in the extra dimension scenario 
\cite{Giddings:2001bu} has been 
proposed. Up to now it remains controversial whether the semi-classical
production cross section is exponentially suppressed
\cite{Voloshin:2001vs} or not 
\cite{Dimopoulos:2001qe}, but even if
there is an exponential suppression the production of mini black holes
at LHC should be still sizably large 
\cite{Rizzo:2002kb}. A couple of semi-classical calculations
have been performed to check the sensitivity of hadron colliders and 
neutrino telescopes to mini-black hole production 
\cite{Bleicher:2001kh}. In this paper we would like to 
address the black hole production from an effective field theoretical
ansatz, constructing an effective interaction Lagrangian and absorbing 
the unknown
quantum physical effects in effective coupling, black hole mass and
the Hawking temperature, hereby
reproducing the semi-classical results in their proper limit. To make
things more definite we would like to study the process:
$e^-(l)+e^-(l') \to {\rm bh} \to  e^-(k)+e^-(k') \;.$

The crucial point is that while the production of the
black hole happens in the vacuum, its decay is a thermal evaporation.
Production and decay are  governed by the same coupling constant, but in
the decay one electron is evaporated thermally and the other is left as a 
remnant. The $e^-e^-$ mode is a possible scenario for future next linear 
colliders like NLC at DESY or CLIC at CERN. We have chosen this process 
because the calculations are quite simple due to the reduced background 
(no strong interactions), but in principle the method is applicable to other 
processes e.~g.~involving hadron colliders as well. 

This technique allows to handle the mini black hole as 
a particle, taking into account quantum physical effects like interference
processes. The absorption of the unknown quantum gravity into couplings,
masses etc. could be seen in parallel to what is done in e.g. effective 
meson field
theory where the unknown collective effects of strong interaction are
absorbed in form factors, effective couplings and masses as well.

The starting point of our consideration 
is the classical black hole thermodynamics because it yields an
explicit expression for the Hawking temperature we are using at the end. 
In analogy to standard thermodynamics one can formulate three basic laws,
see e.~g.~\cite{Traschen:1999zr}:
\begin{itemize}
\item (zeroth law) The surface gravity $\gamma$ of a black hole is constant on
the horizon.
\item (first law) Given $M_{\rm bh}$ the mass, $A$ the area, $L$ the angular
momentum, $\Omega$ the horizon angular velocity, $Q$ the charge and $\nu$
the electrostatic potential (being zero at infinity) of a black hole
one has the energy relation \cite{Bardeen:gs}:
\begin{equation}
\delta M_{\rm bh} = \frac{\gamma}{8\pi} \delta A + \Omega \delta L - 
\nu \delta Q
=T_H \delta S + \Omega \delta L - \nu \delta Q\;.
\end{equation}
\item (second law) the area of a black hole is nondecreasing $\delta A > 0$ 
\cite{Hawking:1971tu}.
\end{itemize}
These laws suggest an identification between the horizon area of a black
hole and its entropy \cite{Bekenstein:ur}, which is given by 
$S_{\rm bh}=A/4$ \cite{Strominger:1996sh}. 
Taking into account the fact that a black hole can evaporate due to the
Hawking radiation \cite{Hawking:sw} one has to complete the area law
by the entropy added by the particle ejected out of the black hole:
\begin{equation}
\delta S_{\rm tot} = \delta S_{\rm outside} + \frac{1}{4} \delta A \ge 0
\;.
\end{equation} 
The different quantities of a black hole can be expressed solely by
the angular momentum $L$, its charge $Q$ and its mass $M_{\rm bh}$. Introducing
the rationalized area $\alpha = A/(4\pi)$ and the orbital angular momentum
parameter $\vec a = \vec L /M_{\rm bh}$ one obtains \cite{Bekenstein:ur}:
\begin{eqnarray}
\alpha &=& r_+^2+a^2 = 2M_{\rm bh}r_+ - Q^2\;, \quad\quad 
\vec \Omega = \frac {\vec a}{\alpha}\,,\nonumber \\
r_\pm  &=& M_{\rm bh}\pm \sqrt{M_{\rm bh}^2-Q^2-a^2}\;, \quad\quad 
 \nu = -\frac{Qr_+}{\alpha}\,,  \nonumber \\
T_{\rm H} &=&\frac{ \gamma}{2\pi} = \frac{(r_+-r_-)}{4\pi\alpha} 
= \frac{2}{A}\sqrt{M_{\rm bh}^2-Q^2-a^2}\,.
\end{eqnarray}
The black hole is then interpreted as a black body radiator with
Hawking temperature $T_H$. 

In this paper we wish to treat the black 
hole as a particle. As such it should be quantized in mass. The mass 
quantization in the conventional four  dimensions has been given in Ref.  
\cite{Beckenstein:1974ab}:
\begin{eqnarray}
M_{{\rm bh}\;n_b q j_z}^2 &=& g M_P^2\left[n_b
\left(1+ \alpha_{\rm em}\frac{q^2}{2n_b}\right)^2 + \frac{j_z^2}{n_b} 
\right]\,,
\nonumber \\
&& j^2_z+\frac{1}{4} \alpha_{\rm em} q^2 \le n_b^2\;,
\end{eqnarray}
where $M_P$ is the Planck mass, q the charge quantum number $j_z$ the angular
momentum quantum number.
The quantum number $n_b$ takes
the quantization of the black hole horizon surface into account
and should not be confused with the number of $n$ extra dimensions used
in later formulas. The pre-factor
$g$ is controversial. In Ref.~\cite{Beckenstein:1974ab} it was
chosen to be $g=1/2$ as the smallest possible quantum unity.
The area quantization has been treated in
the framework of loop quantum gravity (LQG) for the
spheric symmetrical problem, see e.g.~\cite{DePietri:1996pj}. For a special
choice of the quantum numbers for the edges of the surface geometry
one obtains $g=\ln2/(2\pi)$ \cite{Khriplovich:2001je} consistent with
the result derived in \cite{ Ashtekar:1997yu} in the framework of a
Chern Simons field theory. Recently, requiring that the entropy
of the black hole should be maximal a value of $g=0.614/\pi$ has been
derived making use of the LQG result \cite{Khriplovich:2001je}.

\begin{figure}
\includegraphics*[totalheight=5.cm]{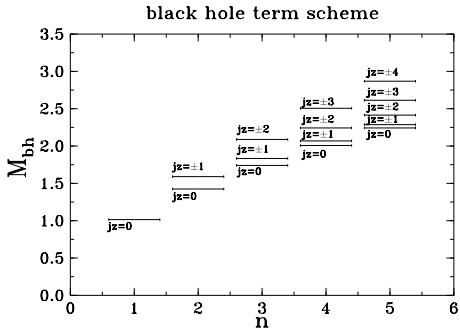}
\caption{Term scheme for a doubly charged black hole $bh^{--}$. The 
black hole mass $M_{\rm bh}$ is given in units of $M_P \sqrt{g}$, 
where $M_P$ is the Planck mass or its equivalent in the extra 
dimension scenario, e.g. $M_P \sim$ 1 TeV.}
\label{termscheme}
\end{figure}

In the extra dimension scenario the fundamental Planck scale $M_P$ should
be at TeV range 
\cite{Arkani-Hamed:1998rs}, which
correspondingly should result in a discrete spectrum for the black holes
spaced in  the order of TeV distances. In $e^-e^- \to bh \to e^-e^-$, 
where we should have a doubly charged
black hole ($q=-2$) we would arrive at the following term scheme, see
Fig.~\ref{termscheme}.  

The black hole quantization predicts that there is an isolated scalar
black hole resonance at $M_P \sqrt{g}$. In the following we concentrate 
on this first scalar excitation  and develop an effective field theory for it. 
In the calculation for the term scheme above we have set 
$\alpha_{\rm em}=1/137$. The running of the electro magnetic coupling
in the conventional standard model as included e.g. in {\sc PYTHIA}
\cite{Sjostrand:2000wi}
up to 10 TeV only makes a difference up to 12\% which
we have neglected here for simplicity, as the effect is not visible here.
It would be a completely different story if also the GUT scale would be
at considerable lower values, but as to a lack of a proper determination
of such a scale we will not pursue this idea further here.
It is our aim to develop a 
workable formalism which allows practical
analysis for high-energy collider reactions. Interferences between
the black hole production and decay and background processes described
by conventional field theory are important. Yet the precise quantum 
gravitational production and evaporation process is not known. 
We therefore want to 
set aside these problems by factorizing the unknown quantum gravity physics
in an effective coupling and regarding only the asymptotic initial and final
states. 

As an example we consider the production of  doubly charged
scalar black holes with angular
momentum j=0 by two fermions e.g. electrons. 
For the interaction part of the Lagrangian one can set:
\begin{eqnarray}
&&{\cal L}_{\rm int} = i\kappa_{\rm eff} M_{\rm bh} \phi \bar \Psi_f  
\hat C \Psi_f 
+ h.c.\,,
\nonumber \\
&&V_{\rm eff}(k_1, k_2) = i\kappa_{\rm eff} M_{\rm bh} \hat C \;,
\end{eqnarray}
where  $\hat C=i\gamma_2 {\cal K}$ is the usual charge conjugation operator
with $\cal K$ being the complex conjugation. 
One should note that $\kappa_{\rm eff}$ may be different for different $n_b$,
so that one has different couplings to different black hole micro states.
The mass scale involved has been chosen to be the
black hole mass $M_{\rm bh}$ and not the fermion mass $m_f$ 
so that the coupling does not
vanish in the limit of vanishing fermion rest mass $m_f$. Such a type
of coupling can be compared to the coupling of two fermions to a
doubly charged Higgs \cite{Barenboim:1996pt}.

As a next step we take an example of 
determining the effective coupling in a crude approximation from the 
production cross section of a black hole by two colliding electrons. 
Here and in the following we will neglect the electron mass $m_e$ throughout, 
because it is more than 6 orders of magnitude
smaller than the TeV scale which sets the black hole mass involved.
Taking the amplitude for black hole production by electrons:
\begin{eqnarray}
&&\hspace*{-5mm}M(p,p')= i\kappa_{\rm eff} M_{\rm bh} \bar u(p) 
\hat C  u(p')\;,
\quad k = \left( M_{\rm bh} ,0,0, 0\right)\;,\nonumber\\
&&\hspace{-5mm} p = \left( \frac{1}{2} \sqrt{s},0,0,\frac{1}{2} 
\sqrt{s} \right)\;,
\quad 
p' = \left( \frac{1}{2} \sqrt{s},0,0,-\frac{1}{2} \sqrt{s} \right),
\end{eqnarray}
one can make a connection with the geometrical production cross section via:
\vspace*{-10mm}
\begin{widetext}
\begin{eqnarray}
\sigma_{\rm bh} &=& \frac{1}{4}
\frac{1}{4 pp'} |M|^2 \int \frac{d^4k}{(2\pi)^3}(2\pi)^4 
 \delta(p+p'-k) \delta\left(k^2-M_{\rm bh}^2\right) 
\nonumber \\
&=&\frac{\pi}{4} \kappa_{\rm eff}^2 M_{\rm bh}  
\delta\left(\sqrt{s}-M_{\rm bh}\right)
\to \pi   \frac{\kappa_{\rm eff}^2}{4} 
\theta\left (\frac{M_{\rm bh}}{2}-|\sqrt{s}-M_{\rm bh}|\right)
=\pi R_S^2 \theta\left (\frac{M_{\rm bh}}{2}-|\sqrt{s}-M_{\rm bh}|\right)
\equiv\sigma_{\rm geom}\;.
\end{eqnarray}
\end{widetext}
In the arrow we transform according to local duality
the $\delta$ function into a finite size step
function of width $M_{\rm bh}$ in order to make contact with the 
geometrical cross section.
$R_{\rm S}$ is the Schwarzschild radius which in $n+3$ dimensions is 
\cite{Myers:un}:
\begin{equation}
R_{\rm S} = \frac{1}{\sqrt{\pi} M_{\rm P}}
\left[\frac{M_{\rm bh}}{M_{\rm P}} 
\left( 
\frac{8 \Gamma\left(\frac{n+3}{2}\right)}{n+2}\right) \right]^{1/(n+1)}\;.
\end{equation}
Here we have suggested $M_{\rm bh}$ as the scale for the effective width
of the black hole production in vacuum
 to compare the field theoretical cross section
to the geometrical one. This comparison then suggests to set 
$\kappa_{\rm eff}/2 = R_{\rm S}$. Of course such a contact between geometrical
and field theoretical cross section is a crude approximation to reality. 
Practically, $\kappa_{\rm eff}$ will  be an effective coupling constant
absorbing the unknown quantum gravitational physics. It is then simply
a constant that has to be determined by experiment. 
%
%
\begin{figure}
\includegraphics*[totalheight=2.cm]{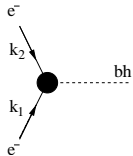}
\caption{Effective vertex for the production process
$e^-e^-\to {\rm bh}$.}
\label{verteff}
\end{figure}
%
%

 As already discussed the evaporation of 
the black hole is thermal. Therefore
it should be possible to describe it 
in the framework of thermal field theory which has been developed
in \cite{Kapusta}. In this connection
we obtain for the partial width bh $\to e^-e^-$ using again 
$k= (M_{\rm bh},0,0,0)$:

\begin{eqnarray}
&&\hspace*{-.7cm}\Gamma_{{\rm bh }\to e^- e^-}=
\frac{\kappa^2_{\rm eff} M_{\rm bh}^2}{2 M_{\rm bh}}
\int\frac{d^4p}{(2\pi)^3} 
\int\frac{d^4p'}{(2\pi)^3}
\frac{1}{e^{\beta_{\rm H}|p_0|}+1}\nonumber \\
&&\hspace*{-.7cm}\times\delta \left(p^2 \right)
\delta \left({p'}^2 \right) 
(2\pi)^4 \delta\left(p+p'-k\right)
{\rm Tr}\left[   p_\mu \gamma^\mu \hat C 
{p'}_\nu \gamma^\nu \hat C^\dagger\right]
\nonumber \\
&&\hspace*{-.7cm}= \frac{\kappa^2_{\rm eff} M_{\rm bh}^3}{8\pi} 
\frac{1}{e^{\beta_{\rm H} M_{\rm bh}/2}+1} \;.
\end{eqnarray}
Here $T_{\rm H}= 1/\beta_{\rm H}$ is the 
Hawking temperature which is in $n+3$  
dimensions given by \cite{Myers:un}:
\begin{equation}
T_{\rm H} = M_{\rm P} \left( \frac{M_{\rm P}}{M_{\rm bh}} 
\frac{n+2}{8 \Gamma\left( \frac{n+3}{2} \right)}\right)^{1/(n+1)}
\frac{n+1}{4\sqrt{\pi}}\;.
\end{equation}
One should note that the factor $1/(\exp(\beta_{\rm H} M_{\rm bh}/2)+1)$
does not belong to the coupling but to the outgoing electrons. The principle
should be that each evaporated particle goes with the corresponding thermal
occupation number due to the proper statistic it belongs to. For the final
evaporation it is actually only one particle that evaporates while the other
one is just the remaining remnant where the 
energy is fixed by energy momentum conservation. 
As one can not distinguish in our case
which of the electrons has evaporated and which is the remnant we end
up with a single factor $1/(\exp(\beta_{\rm H} M_{\rm bh}/2)+1)$. If we had
a different black hole evaporating in an electron and a photon we had
to multiply with a factor 
%
%
$\left(1/(e^{\beta_{\rm H} M_{\rm bh}/2}-1)+
       1/(e^{\beta_{\rm H} M_{\rm bh}/2}+1)\right)
$
to allow one time for a photon and one time for an electron evaporation
whereas the other particle is just the remnant.

\begin{figure}
\includegraphics*[totalheight=5.cm]{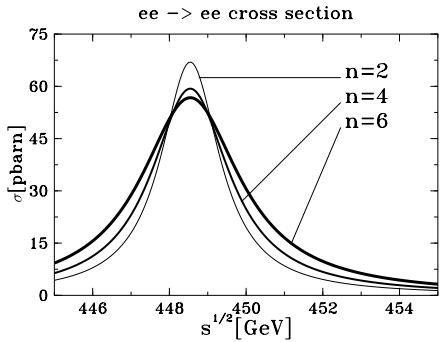}
\caption{Total cross section for the reaction $ee\to {\rm bh} \to ee$ in the
Khriplovich scenario \cite{Khriplovich:2001je}.}
\label{sigma}
\end{figure}

As a last application of the 
method we calculate the $e^-e^- \to e^- e^-$ cross section, for the first 
scalar black hole excitation using the method of effective thermal field 
theory discussed above. For the amplitude we find c.f. Fig.~\ref{eeeegraph}:
%
%
\begin{figure}
\includegraphics*[totalheight=2.cm]{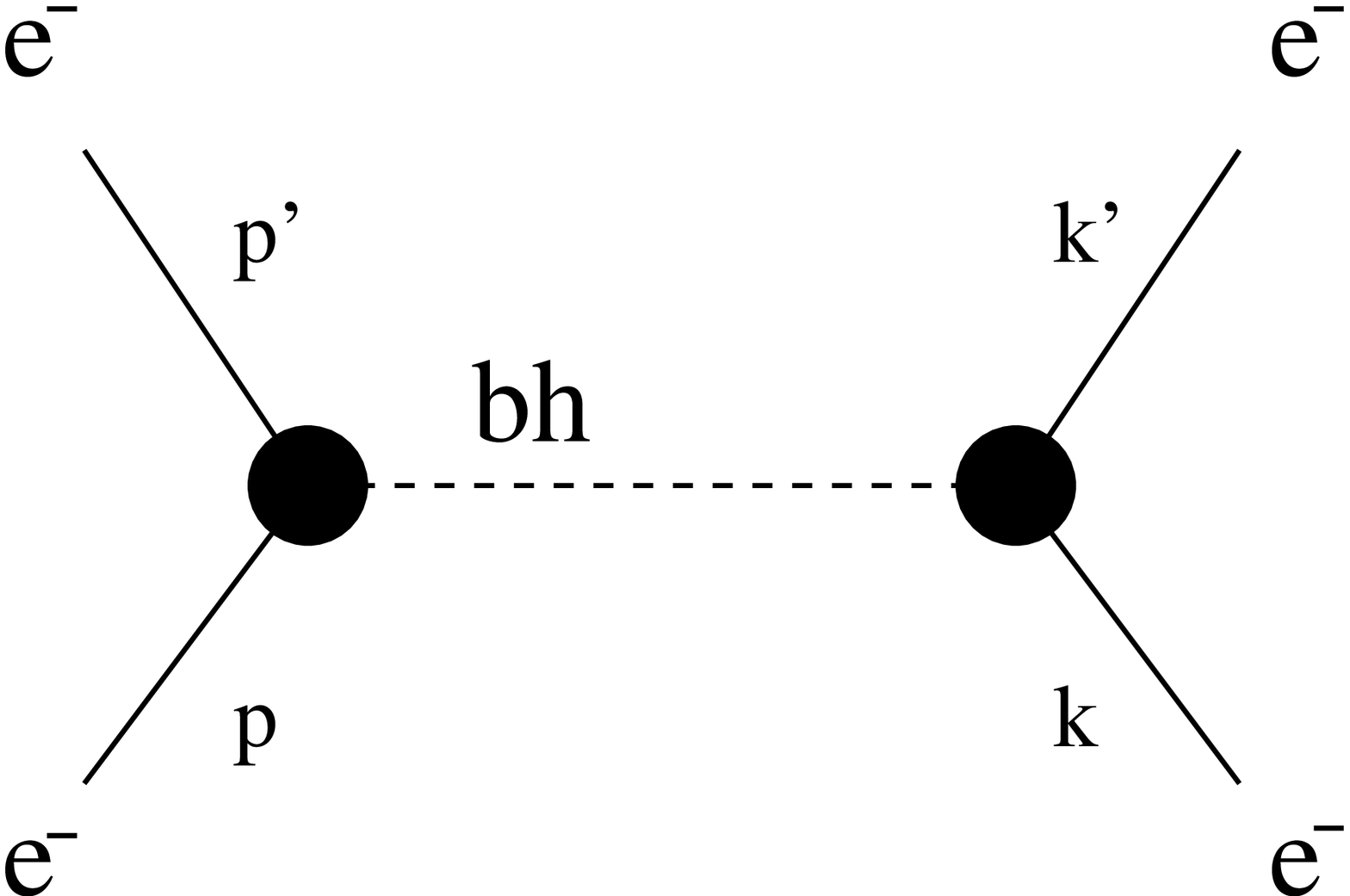}
\caption{Graph for the process $e^-e^-\to {\rm bh} \to e^-e^-$.}
\label{eeeegraph}
\end{figure}
%
%
\begin{eqnarray}
{\cal M} = M(p,p') 
\frac{1}{s-\left(M_{\rm bh}+
 \frac{i}{2}\Gamma_{
{\rm total}}\right)^2}
M(k,k').
\end{eqnarray}
\noindent 
Then the corresponding cross section has the form:
\vspace*{-5mm}
\begin{widetext}
\begin{eqnarray}
\sigma(s) &=& \frac{1}{4} 
\int \frac{d^4 k }{(2\pi)^3} \delta\left(k^2\right)
\int\frac{d^4 k'}{(2\pi)^3} \delta\left(k'^2\right)
\frac{|{\cal M}|^2}{4 pp'}
\frac{1}{e^{\beta_{\rm H} k_0}+1}
(2\pi)^4 \delta\left(p+p'-k-k'\right)
\nonumber \\
&=& \frac{\kappa_{\rm eff}^4 M_{\rm bh}^4 s}{16\pi} \frac{1}
{\left(s-\left(M_{\rm bh}^2 - \frac{\Gamma_{{\rm total}}^2}{4} 
\right)\right)^2
+ M_{\rm bh}^2 \Gamma_{{\rm total}}^2} 
\frac{1}{e^{\beta_{\rm H} \sqrt{s}/2}+1}\;.
\end{eqnarray}
\end{widetext}
In Fig.~\ref{sigma} we show the total cross section $ee\to {\rm bh} \to ee$
in the Khriplovich scenario Ref.~\cite{Khriplovich:2001je}, i.e.~for
a first scalar black hole mass of $M_{\rm bh}\approx 448.5\;{\rm GeV}$ assuming
$M_{\rm P} = 1\;{\rm TeV}$ for different values of n (extra dimensions).
One should keep in mind that at such comparatively small
energies, i.e. smaller than a few times
the Planck mass, it may be controversial whether such a resonance should
be interpreted as a genuine black hole or rather as a string excitation.
We have set to leading order  $\Gamma_{\rm total} = 
\Gamma_{{\rm bh}\to e^-e^-}$ assuming
strict lepton number conservation, 
as current bounds from muon decay into three electrons otherwise 
require a Planck scale in the 100 TeV range.
It is seen that the width of the resonance in first order 
is about 2-3 GeV and that the total cross section is in the range of pbarn.
%

In this paper we intend to give a 
description for the black hole in terms of a particle. Here the mini black 
holes  should arise as discrete resonances in accordance with the idea of 
Beckenstein that the black hole area should be quantized.
The in vacuum production of black holes can be described by means of standard
field theory where the unknown quantum gravity is absorbed in an 
effective coupling constant $\kappa_{\rm eff}$. The thermal evaporation 
is correspondingly described in terms of the thermal field theory.
Such a description allows to take into consideration quantum interference
effects which are not accessible in a semi-classical description and will
be helpful for the experimental analysis of events that may come from
mini-black holes produced at high energy scattering experiments.
The formalism shown here  can be easily generalized
to a variety of processes,
for example also to mini black hole production at LHC. Within this
framework and restricting to the well
isolated first scalar black hole resonance,
the analysis of the quantum gravitational effects boils down
to the measurement of only three independent quantities: 
The effective coupling $\kappa_{eff}$,  $M_{\rm bh}$ (the first
scalar black hole mass) and the Hawking temperature 
$T_{\rm H}=1/\beta_{\rm H}$. 
In this way studying definite subprocesses in possible black hole production
in terms of an effective quantum field theory may allow to pin down the
quantum gravitational content to actually three parameters. These parameters
as soon as obtained experimentally can then be compared to various 
quantum gravitational scenarios to improve our understanding
of the fundamental laws of gravitational physics. 

Acknowledgment: We wish to acknowledge fruitful and stimulating discussion
with J.~Bijnens, S.~Giddings, G.~Gustafson and T.~Sj\"ostrand. 
This work was supported in part by the Swedish Research Council.


\begin{thebibliography}{99}
\bibitem{Giddings:2001bu}
S.~B.~Giddings and S.~Thomas,
Phys.\ Rev.\ D {\bf 65} (2002) 056010
[arXiv:hep-ph/0106219];
S.~Dimopoulos and G.~Landsberg,
Phys.\ Rev.\ Lett.\  {\bf 87} (2001) 161602
[arXiv:hep-ph/0106295];
G.~Landsberg,
arXiv:hep-ph/0112061.



\bibitem{Voloshin:2001vs}
M.~B.~Voloshin,
Phys.\ Lett.\ B {\bf 518} (2001) 137
[arXiv:hep-ph/0107119];
M.~B.~Voloshin,
Phys.\ Lett.\ B {\bf 524} (2002) 376
[arXiv:hep-ph/0111099].



\bibitem{Dimopoulos:2001qe}
S.~Dimopoulos and R.~Emparan,
Phys.\ Lett.\ B {\bf 526} (2002) 393
[arXiv:hep-ph/0108060];
S.~N.~Solodukhin,
arXiv:hep-ph/0201248;
D.~M.~Eardley and S.~B.~Giddings,
arXiv:gr-qc/0201034.






\bibitem{Rizzo:2002kb}
T.~G.~Rizzo,
JHEP {\bf 0202} (2002) 011
[arXiv:hep-ph/0201228];
T.~G.~Rizzo,
in {\it Proc. of the APS/DPF/DPB Summer Study on the Future of Particle 
Physics (Snowmass 2001) } ed. R.~Davidson and C.~Quigg,
arXiv:hep-ph/0111230.


\bibitem{Bleicher:2001kh}
M.~Bleicher, S.~Hofmann, S.~Hossenfelder and H.~St\"ocker,
arXiv:hep-ph/0112186;
S.~Hofmann, M.~Bleicher, L.~Gerland, S.~Hossenfelder, S.~Schwabe and 
H.~St\"ocker, arXiv:hep-ph/0111052;
S.~Hossenfelder, S.~Hofmann, M.~Bleicher and H.~St\"ocker,
arXiv:hep-ph/0109085;
M.~Kowalski, A.~Ringwald and H.~Tu,
Phys.\ Lett.\ B {\bf 529} (2002) 1
[arXiv:hep-ph/0201139];
A.~Ringwald and H.~Tu,
Phys.\ Lett.\ B {\bf 525} (2002) 135
[arXiv:hep-ph/0111042].






\bibitem{Khriplovich:2001je}
I.~B.~Khriplovich,
arXiv:gr-qc/0109092.



\bibitem{Traschen:1999zr}
J.~Traschen,
arXiv:gr-qc/0010055.


\bibitem{Bardeen:gs}
J.~M.~Bardeen, B.~Carter and S.~W.~Hawking,
Commun.\ Math.\ Phys.\  {\bf 31} (1973) 161.




\bibitem{Hawking:1971tu}
S.~W.~Hawking,
Phys.\ Rev.\ Lett.\  {\bf 26} (1971) 1344.



\bibitem{Bekenstein:ur}
J.~D.~Bekenstein,
Phys.\ Rev.\ D {\bf 7} (1973) 2333.





\bibitem{Strominger:1996sh}
A.~Strominger and C.~Vafa,
Phys.\ Lett.\ B {\bf 379} (1996) 99
[arXiv:hep-th/9601029].





\bibitem{Hawking:sw}
S.~W.~Hawking,
Commun.\ Math.\ Phys.\  {\bf 43} (1975) 199.




\bibitem{Beckenstein:1974ab}
J.~D.~Beckenstein
Lett.\ Nuov.\ Cim.\ {\bf 11} (1974) 467.



\bibitem{DePietri:1996pj}
R.~De Pietri and C.~Rovelli,
Phys.\ Rev.\ D {\bf 54} (1996) 2664
[arXiv:gr-qc/9602023].



\bibitem{Ashtekar:1997yu}
A.~Ashtekar, J.~Baez, A.~Corichi and K.~Krasnov,
Phys.\ Rev.\ Lett.\  {\bf 80} (1998) 904
[arXiv:gr-qc/9710007].

\bibitem{Arkani-Hamed:1998rs}
N.~Arkani-Hamed, S.~Dimopoulos and G.~R.~Dvali,
Phys.\ Lett.\ B {\bf 429} (1998) 263
[arXiv:hep-ph/9803315];
I.~Antoniadis, N.~Arkani-Hamed, S.~Dimopoulos and G.~R.~Dvali,
Phys.\ Lett.\ B {\bf 436} (1998) 257
[arXiv:hep-ph/9804398];
N.~Arkani-Hamed, S.~Dimopoulos and G.~R.~Dvali,
Phys.\ Rev.\ D {\bf 59} (1999) 086004
[arXiv:hep-ph/9807344].


\bibitem{Sjostrand:2000wi}
T.~Sjostrand, P.~Eden, C.~Friberg, L.~Lonnblad, G.~Miu, S.~Mrenna and 
E.~Norrbin, Comput.\ Phys.\ Commun.\  {\bf 135} (2001) 238
[arXiv:hep-ph/0010017].



%
%

\bibitem{Barenboim:1996pt}
G.~Barenboim, K.~Huitu, J.~Maalampi and M.~Raidal,
Phys.\ Lett.\ B {\bf 394} (1997) 132
[arXiv:hep-ph/9611362].



\bibitem{Myers:un}
R.~C.~Myers and M.~J.~Perry,
Annals Phys.\  {\bf 172} (1986) 304.

\bibitem{Kapusta}
J.~I.~Kapusta, `` Finite-temperature field theory'', Cambridge
University Press 1989.


\end{thebibliography}
\end{document}